\documentclass[review]{elsarticle}
\pdfoutput=1
\usepackage{lineno,hyperref}
\usepackage{ graphicx, rotating }
\usepackage{ amssymb }
\usepackage{epstopdf}
\usepackage{color}

\journal{Nuclear Instruments and Methods A}

\def\zBB{$0\nu\beta\beta$}

\def\BB{$\beta\beta$}

\def\Tz{$T^{0\nu}_{1/2}$}

\def\qval{$Q_{\beta\beta}$}                 
\def\Gerda{{\sc Gerda}}             
\def\MJ{{\sc Majorana}}             
\def\DEM{{\sc Demonstrator}}             
\def\be{\begin{equation}}
\def\ee{\end{equation}}

\def\gess{$^{76}\mathrm{Ge}$}

\begin{document}

\begin{frontmatter}

\title{The Processing of Enriched Germanium for the \MJ\ \DEM\ and R\&D for a Possible Future Ton-Scale \gess\ Double-Beta Decay Experiment}

\author[lbnl]{N.~Abgrall}		
\author[pnnl]{I.J.~Arnquist}
\author[usc,ornl]{F.T.~Avignone~III\corref{cor}}
\ead{avignone@physics.sc.edu}
\author[ITEP]{A.S.~Barabash}	
\author[ornl]{F.E.~Bertrand}
\author[lbnl]{A.W.~Bradley}
\author[JINR]{V.~Brudanin}
\author[duke,tunl]{M.~Busch}	
\author[uw]{M.~Buuck}
\author[ecs]{J.~Caja\fnref{deceased}}
\author[ecs]{M.~Caja}
\author[unc,tunl]{T.S.~Caldwell}
\author[sdsmt]{C.D.~Christofferson}
\author[lanl]{P.-H.~Chu}
\author[uw]{C. Cuesta}
\author[uw]{J.A.~Detwiler}
\author[sdsmt]{C.~Dunagan}
\author[ecs]{D.T.~Dunstan}
\author[ut]{Yu.~Efremenko}
\author[ou]{H.~Ejiri}
\author[lanl]{S.R.~Elliott}
\author[unc,tunl]{T.~Gilliss}
\author[princeton]{G.K.~Giovanetti}
\author[lanl]{J.~Goett}	
\author[ncsu,tunl,ornl]{M.P.~Green}
\author[uw]{J.~Gruszko}
\author[uw]{I.S.~Guinn}		
\author[usc]{V.E.~Guiseppe}
\author[unc,tunl]{C.R.S.~Haufe}
\author[unc,tunl]{R.~Henning}
\author[pnnl]{E.W.~Hoppe}
\author[usd]{B.R.~Jasinski}
\author[ttu]{M.F.~Kidd}	
\author[ITEP]{S.I.~Konovalov}
\author[pnnl]{R.T.~Kouzes}
\author[ut]{A.M.~Lopez}
\author[unc,tunl]{J.~MacMullin}
\author[queens]{R.D.~Martin}
\author[lanl]{R. Massarczyk}
\author[unc,tunl]{S.J.~Meijer}	
\author[mpi,tum]{S.~Mertens}
\author[meyer]{J.H.~Meyer}
\author[lbnl]{J.~Myslik}		
\author[unc,tunl]{C.~O'Shaughnessy}	
\author[lbnl]{A.W.P.~Poon}
\author[ornl]{D.C.~Radford}
\author[unc,tunl]{J.~Rager}
\author[unc,tunl]{A.L.~Reine}
\author[meyer]{J.A.~Reising}	
\author[lanl]{K.~Rielage}
\author[uw]{R.G.H.~Robertson}
\author[unc,tunl]{B.~Shanks}	
\author[JINR]{M.~Shirchenko}
\author[sdsmt]{A.M.~Suriano}
\author[usc]{D.~Tedeschi}
\author[ecs]{L.M.~Toth}
\author[unc,tunl]{J.E.~Trimble}
\author[ornl]{R.L.~Varner}
\author[JINR]{S.~Vasilyev}
\author[lbnl]{K.~Vetter\fnref{ucb}}
\author[unc,tunl]{K.~Vorren}
\author[lanl]{B.R.~White}	
\author[unc,tunl,ornl]{J.F.~Wilkerson}
\author[usc]{C.~Wiseman}		
\author[unc,tunl]{W.~Xu}
\author[JINR]{E.~Yakushev}
\author[ornl]{C.-H.~Yu}
\author[ITEP]{V.~Yumatov}
\author[JINR]{I.~Zhitnikov}
\author[lanl]{B.X.~Zhu}

\address[lbnl]{Nuclear Science Division, Lawrence Berkeley National Laboratory, Berkeley, CA, USA} 
\address[pnnl]{Pacific Northwest National Laboratory, Richland, WA, USA}
\address[usc]{Department of Physics and Astronomy, University of South Carolina, Columbia, SC, USA}
\address[ornl]{Oak Ridge National Laboratory, Oak Ridge, TN, USA} 
\address[ITEP]{National Research Center ``Kurchatov Institute'' Institute for Theoretical and Experimental Physics,
Moscow, Russia} 
\address[JINR]{Joint Institute for Nuclear Research, Dubna, Russia}
\address[duke]{Department of Physics, Duke University, Durham, NC, USA}
\address[tunl]{Triangle Universities Nuclear Laboratory, Durham, NC, USA}
\address[uw]{Center for Experimental Nuclear Physics and Astrophysics, and Department of Physics, University of
Washington, Seattle, WA, USA}
\address[ecs]{Electrochemical Systems Inc., Oak Ridge, Tennessee, USA}
\address[unc]{Department of Physics and Astronomy, University of North Carolina, Chapel Hill, NC, USA}
\address[sdsmt]{South Dakota School of Mines and Technology, Rapid City, SD, USA}
\address[lanl]{Los Alamos National Laboratory, Los Alamos, NM, USA} 
\address[ut]{Department of Physics and Astronomy, University of Tennessee, Knoxville, TN, USA} 
\address[ou]{Research Center for Nuclear Physics and Department of Physics, Osaka University, Ibaraki, Osaka, Japan}
\address[princeton]{Department of Physics, Princeton University, Princeton, NJ, USA}
\address[ncsu]{Department of Physics, North Carolina State University, Raleigh, NC, USA} 
\address[usd]{Department of Physics, University of South Dakota, Vermillion, SD, USA} 
\address[ttu]{Tennessee Tech University, Cookeville, TN, USA} 
\address[queens]{Department of Physics, Engineering Physics and Astronomy, Queen's University, Kingston, ON, Canada} 
\address[mpi]{Max-Planck-Institut f\"{u}r Physik, M\"{u}nchen, Germany}
\address[tum]{Physik Department and Excellence Cluster Universe, Technische Universit\"{a}t, M\"{u}nchen, Germany}
\address[meyer]{J.H. Meyer Consultants LLC, Plainfield, IN}

\fntext[deceased]{Deceased}
\fntext[ucb]{Alternate Address: Department of Nuclear Engineering, University of California, Berkeley, CA, USA}
\cortext[cor]{Corresponding author}

\begin{abstract}
The \MJ\ \DEM\ is an array of point-contact Ge detectors fabricated from Ge isotopically enriched to 88\% in \gess\ to search for neutrinoless double beta decay. The processing of Ge for germanium detectors is a well-known technology. However, because of the high cost of Ge enriched in \gess\, special procedures were required to maximize the yield of detector mass and to minimize exposure to cosmic rays. These procedures include careful accounting for the material; shielding it to reduce cosmogenic generation of radioactive isotopes; and development of special reprocessing techniques for contaminated solid germanium, shavings, grindings, acid etchant and cutting fluids from detector fabrication. Processing procedures were developed that resulted in a total yield in detector mass of 70\%. However, none of the acid-etch solution and only 50\% of the cutting fluids from detector fabrication were reprocessed. Had they been processed, the projections for the recovery yield would be between 80 -- 85\%. Maximizing yield is critical to justify a possible future ton-scale experiment. A process for recovery of germanium from the acid-etch solution was developed with yield of  about 90\%. All material was shielded or stored underground whenever possible to minimize the formation of $^{68}$Ge by cosmic rays, which contributes background in the double-beta decay region of interest and cannot be removed by zone refinement and crystal growth. Formation of $^{68}$Ge was reduced by a significant factor over that in natural abundance detectors not protected from cosmic rays. 
\end{abstract}


\end{frontmatter}


\section{Introduction}

Nuclear double-beta decay without emission of neutrinos, zero-neutrino double-beta decay (\zBB\ decay), is of great current interest in fundamental physics. First, the decay violates conservation of lepton number. Second, it is the only practical way to determine whether neutrinos are their own antiparticles, i.e., Majorana particles. Third, in the case of the light neutrino exchange mechanism, if \zBB\ decay is observed and the half-life measured and combined with neutrino oscillation data, the masses of all three neutrino-mass eigenstates would be determined; hence the neutrino mass scale. There are many recent reviews covering both theoretical and experimental aspects of this subject \cite{Ell00, Ell04, Avi08, Bar11, Gui12}. 
The probability of a direct observation is enhanced by the parameters obtained by the measurements of neutrino oscillations of solar 
 \cite{Ahm01}, atmospheric \cite{Fuk98}, accelerator \cite{Eva13, Ahn06} and reactor neutrinos \cite{Mar12, Egu03, Ahn12, An12, Abe12}.

The importance of low-energy neutrino physics was clearly demonstrated by the awarding of the 2015 Nobel Prize in Physics  jointly to Takaaki Kajita and Arthur B. McDonald ``for the discovery of neutrino oscillations" \cite{Nobel15}. The basis of the prize was the discovery of oscillation of atmospheric neutrinos \cite{Kaj10}, the direct measurement of solar neutrino flavor transformation \cite{Bel16}, and direct evidence that the flux of $^8$B neutrinos from the sun, predicted by Bahcall and his colleagues, is correct \cite{Bah03}. 
Under the assumption that the  \zBB\ decay process is driven by a massive neutrino exchange, the directly measured neutrino oscillation parameters imply that the Majorana mass of the electron neutrino could be larger than 50 meV. This is a scale at which \zBB\ decay might be observed by a ton-scale experiment. Recently, the Capozzi et al.~\cite{Cap17} review article gives references to all of the oscillation results and provides a global fit to all of the neutrino-oscillation data. 
A large, perhaps ton-scale, \zBB-decay experiment might well be motivated by the results presented in that review.
Further, the U.S. Nuclear Science Advisory Committee's  2015 Long Range Plan \cite{nsac15} recommends  ``... \emph{the timely development and
deployment of a U.S.-led ton-scale neutrinoless
double-beta decay experiment.}''

The \MJ\ \DEM\ is a research and development project to determine if a ton-scale \gess\ \zBB-decay experiment is feasible. Two main requirements to demonstrate such feasibility are the highest possible yield in the total mass converted into detectors with the lowest possible radioactive background. Due to the high cost of enriched \gess, processing of the material from an oxide to a high-resistivity, detector-grade metal\footnote{Throughout this paper, the product of GeO$_2$ reduction is referred to as a metal to stay consistent with industry jargon, when in fact the Ge is an electrical semiconductor or metalloid.} must provide the highest possible yield. In addition, high-yield reprocessing of the ``scrap'' germanium from detector fabrication, as well as the efficient recovery of enriched germanium from the acid etch solution, and from grindings and shavings mixed with cutting fluids, herein called sludge, must be achieved. In this article, we describe the techniques developed and used in processing enriched germanium for the \MJ\ \DEM. 

\section{Double-Beta Decay Experiments}

The germanium detector is a well-known device with many applications from the detection of gamma rays from nuclear reactions to environmental radiological evaluations. Germanium detectors were first introduced into the field of neutrino physics by Ettore Fiorini and his colleagues, in their first search for the  decay of \gess\ to $^{76}$Se \cite{Fio67}. This experiment used the concept of the detector and the source of the decay isotope being one and the same, resulting in large detection efficiency. Until recently, the most sensitive searches for neutrinoless double-beta decay came from the first two experiments utilizing detectors fabricated from germanium enriched to 86\% in $^{76}$Ge from the natural abundance of 7.8\%. They were the IGEX experiment:  \Tz$(^{76}Ge)>1.6\times10^{25}$ y \cite{Aal99, Aal02, Aal04} and the Heidelberg-Moscow experiment: \Tz$(^{76}Ge) >1.9\times10^{25}$ y \cite{Kla01}.  Recently, data from CUORE-0, the TeO$_2$ bolometer search for the \zBB\ decay of $^{130}$Te \cite{Alf15}, was combined with those from the CUORICINO experiment \cite{Arn08}, to yield a lower limit: \Tz$(^{130}Te)>4.0\times10^{24}$ y.  There are also recent results from the EXO-200 experiment: \Tz$(^{136}Xe)>1.1\times10^{25}$ y \cite{Alb14} and from the KamLAND-Zen experiment: \Tz$(^{136}Xe)>1.07\times10^{26}$ y \cite{Gan16}. These experiments have claimed stronger upper bounds on $m_{\beta\beta}$, the effective Majorana mass of the electron neutrino; however, large uncertainties in the nuclear matrix elements used to determine bounds on the $m_{\beta\beta}$ from the half-life limits make a clear distinction difficult. Nevertheless, the Kamland-Zen experiment bound implies  an upper limit of 61-165 meV, using the most least favorable published nuclear matrix elements \cite{Gan16}.

The \Gerda\ Phase-II experiment is an array of Broad-Energy Germanium (BEGe) detectors suspended in liquid argon, which cools the detectors and also acts as an active veto detector to cancel background events \cite{Ago14,Ago15}. \Gerda\ is operating in the Laboratori Nazionali del Gran Sasso (LNGS) in Assergi, Italy. The \Gerda\ Phase-I experiment used the semi-coaxial enriched germanium detectors from the Heidelberg-Moscow and the IGEX experiments. The detector collected 21.6 kg-y of data and set a lower limit:  \Tz$(^{76}Ge)>2.1\times10^{25}$~y (90\% C.L.) while when combined with the Heidelberg-Moscow and IGEX data results in the lower limit: \Tz$(^{76}Ge)>3.0\times10^{25}$~y (90\% C.L.) \cite{Ago13}.  This result was used in an attempt to exclude the claim of discovery by H.V. Klapdor-Kleingrothaus et al. \cite{Kla01a, Kla04a, Kla04b}. A recent joint analysis of the  \Gerda\  Phase I and II sets a lower limit of  \Tz$(^{76}Ge)>5.3\times10^{25}$~y (90\% C.L.) \cite{Ago17}.

\section{The \uppercase{M\footnotesize{ajorana} \normalsize{D}\footnotesize{emonstrator}}}

\begin{figure}[h]
\centering
\includegraphics
[width=0.8\textwidth]{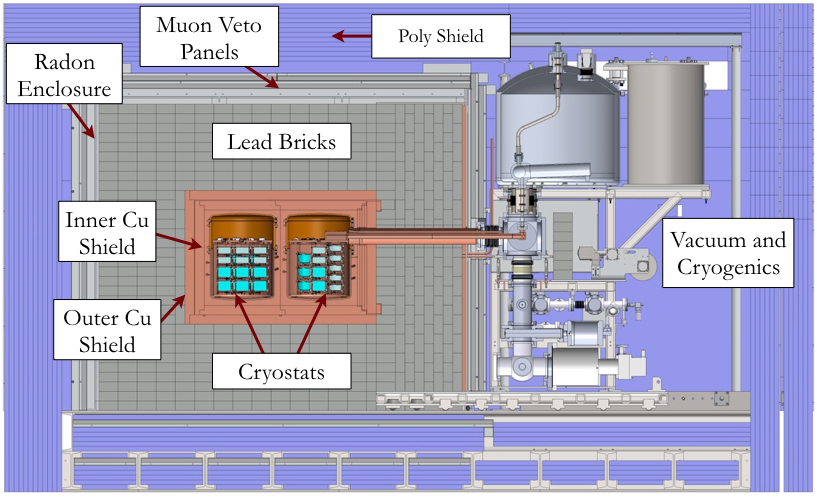}
\caption[]{A computer-aided design drawing of the \MJ\ \DEM\ showing the detector arrays, cooling system, and complex shielding of copper, lead, and a 4$\pi$ cosmic-muon active veto surrounded by polyethylene neutron-moderating shield.    The inner polyethylene panels are borated to absorb moderated neutrons.}  
\label{fig:MJD}
\end{figure}

The \MJ\ \DEM\  \cite{Abg14} is composed of two arrays of point-contact germanium detectors~\cite{Luk86} in a common shield as shown in Figure \ref{fig:MJD}. The two arrays contain a total of 29.7 kg of detectors fabricated from germanium enriched to 88\% in the \BB-decay isotope \gess.   The arrays also contain 14.4 kg of detectors of natural abundance germanium. The average enriched detector mass is 850 g. 

The experiment is operating at a depth of 4850 ft below the surface in the Sanford Underground Research Facility (SURF) in the former Homestake Gold Mine in Lead, South Dakota. This site has 4260 meters of water equivalent overburden~\cite{Hei15} to shield the experiment from cosmic-ray muons and the high-energy neutrons they generate. The main advantage of point-contact germanium detectors is their excellent pulse-shape discrimination capability; this allows the identification and removal from the spectrum of multi-site events typical of MeV gamma-ray backgrounds. In addition, the low-energy threshold ($<$500 eV) that can be achieved using p-type point contact detectors opens up new physics programs for the \MJ\ \DEM, including detection of light WIMP dark matter and solar axions interacting with the detector~\cite{Abg16}. The reduction of cosmogenic activation is essential to maintain this low-energy physics program. See Section \ref{sec:cosmo} for a discussion of the cosmogenic generation of radioactive isotopes. At every step of the germanium processing, precautions were taken to minimize this source of background.

A detailed discussion of the properties of the detectors and their pulse shape discrimination capabilities is given in reference \cite{Abg14}. Thirty-five point-contact detectors have been produced from the enriched material, comprising 29.7 kg of germanium enriched to 88\% in \gess. Twenty of these detectors are operating in Module 1 of the \MJ\ \DEM\ with the remaining deployed in Module 2. The remainder of this article provides a brief discussion of the acquisition and reduction of the enriched \gess O$_2$, the zone refinement of the metal to a resistivity of $\geq 47\ \Omega\cdot$cm, and the reprocessing of the scrap material from detector fabrication and liquids returned from the detector manufacturer. 

\section{Germanium Detector Fabrication}

It is appropriate to briefly discuss the processes involved in the fabrication of germanium detectors that have important impacts on the reprocessing and conservation of the enriched germanium. The production of point-contact detectors was carried out by AMETEK-ORTEC Inc. at their facility in Oak Ridge, Tennessee located near the \MJ\ \DEM\ germanium processing facility. The fabrication of point-contact detectors presented new challenges. The detectors used in the IGEX, Heidelberg-Moscow, and subsequently the \Gerda\ Phase-I experiment, were large (2 kg) semi-coaxial detectors for which there were many years of experience in the industry \cite{Kno89}. The \Gerda\ Phase-II and \MJ\ \DEM\ use the new technologies of point-contact style germanium detectors, which provide much improved discrimination between single-site interactions similar to double-beta decay interactions and multi-site events characteristic of gamma-ray background \cite{Luk86, Bar07}. The fabrication of germanium detectors requires etching the surfaces of the detector crystal blanks 
at various stages of the production, resulting in about a 2\% loss of germanium per etch. These losses must be minimized when producing detectors from isotopically enriched material.

Detector manufacturers require the input material to be in the form of zone-refined germanium bars with resistivity levels of 47~$\Omega\cdot$cm or higher, or equivalently 10$^{13}$ electrically active impurities/cm$^3$. The detector fabricator zone-refines the material again to a level of about 10$^{11}$ or lower in the same units. The metal is then introduced into the crucible of a Czochralski crystal puller. The temperature is raised to liquefy the germanium, and a seed crystal is introduced and pulled very slowly to form a crystal boule from which the detector blank is machined. The crystal-pulling process further purifies the germanium metal. The blank is etched with nitric and hydrofluoric (HF) acid solutions before all but one surface of the blank is diffused with lithium to form the p-n junction of the p-type germanium semi-conductor diode.

These steps produce a significant quantity of valuable ``scrap'' germanium that must be recovered. The acid etch can contain as much as 7--10\% of the original mass of the germanium blank. The machining scraps (kerf) and tails cut from the boule can represent several kilograms of materiel. Frequently, half of the tails can be  used by the detector fabricator with no further processing. The remainder must be recovered in the case of enriched germanium. In the next section we discuss the processing, reprocessing, and recovery of scrap material used in the \MJ\ \DEM.

\section{Special Germanium Processing Procedures for the \uppercase{M\footnotesize{ajorana} \normalsize{D}\footnotesize{emonstrator}}} 

As noted above, the objective of the germanium processing described in this paper was to provide \gess\ suitable for fabrication of germanium detectors for the \MJ\ \DEM. The processing described in this section involves several components: material enrichment and shipping, reduction of the material from oxide to metal, zone refinement of the germanium metal into bars having the appropriate resistivity for detector manufacture, and reprocessing of material for reuse in detector fabrication. The latter category includes reprocessing of metal pieces returned from the detector manufacturer, liquid ``sludge'' (consisting of water and cutting fluid that contain enriched germanium), 
and the acid from etching the metal at various stages of the process of detector fabrication. 
The materials requiring reprocessing can represent a significant fraction of the total mass of germanium being processed.  Detector manufacturers usually discard the acid etch and sludge. However, the high cost of enriched germanium required the development of cost-effective methods to recover germanium suspended in the liquids. A special facility was set up by the collaboration in Oak Ridge, Tennessee and managed by Electrochemical Systems, Inc. (ESI) to carry out the reduction of the \gess O$_2$, zone refining of the reduced metal, and recovery  of the scrap germanium. 

\begin{figure}[h]
\centering
\includegraphics
[width=0.8\textwidth]{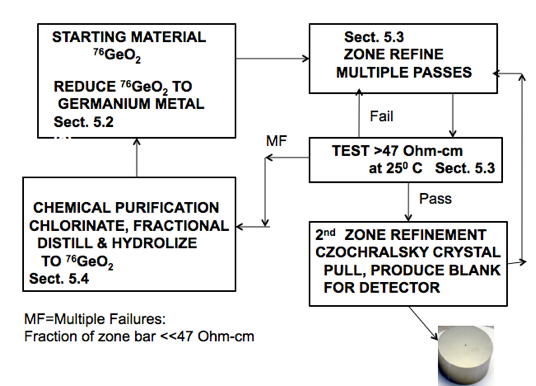}
\caption[]{A diagram of the purification process used for the \MJ\ \DEM\ germanium enriched to 88\% in \gess . The arrow labeled ``Multiple Failures'' represents a very small fraction of the original 42.5~kg of Ge. The section where the process is explained is noted in each block.}
\label{fig:flow}
\end{figure}

The process of reduction of the GeO$_2$ and purification of the metal by multiple zone refinement is shown in the flow chart in Figure \ref{fig:flow}. It differs significantly from the process used in the germanium industry. In the process described here, the metal from reduction is all sent to zone refinement. The resistivity of the resulting zone-refined metal bars is measured with a 3-point resistivity probe. The portions that measure above $47\ \Omega\cdot$cm are cut off and sent to the detector manufacturer. The rest is returned to the zone refiner with newly reduced metal. This process is repeated many times until there is not enough germanium to continue the process. 

In the typical industrial process, the metal bars from the reduction furnace are measured and the portion that measures below 5 $\Omega\cdot$cm is chemically purified by chlorination, as described later. The chlorination chemical purification has an average yield of 70\%, while the rezoning technique used in the present work has almost no loss. Accordingly, the processing of all of the germanium in this project had a yield of 98.3\% for the reduction and zone refinement of all of the virgin material (see Table~\ref{table:ge_load}). 

In a standard detector fabrication run at the manufacturer (AMETEK-ORTEC), a charge of approximately 9~kg of $^{76}$Ge-enriched metal was zone refined further before loading into a crystal puller.  Each of these charges would contain a combination of input $^{76}$Ge metal that had different processing histories.  In the initial phase of detector fabrication, these charges were predominantly virgin electronic-grade metals that ESI produced directly from the reduction of $^{76}$GeO$_2$ and initial zone refinement.  During the intermediate phase of detector fabrication, the proportion of `recycled' material --- unused enriched metal from previously pulled crystals that could be reloaded in the zone refiner after an acid etch at the detector manufacturer --- in each charge would increase.  As more detectors were manufactured and more potentially recoverable enriched materials (kurf and sludge from cutting) were collected and chemically reprocessed by our ESI team, the amount of these `recoverable' materials would increase in the later stage of detector fabrication.  Table~\ref{table:ge_load} is a summary of the composition of these three different input streams of metals for the charges in each of the thirteen zone-refined bars in detector fabrication.  The summed mass of detectors that were produced from each zone-refined bar is also shown in this table, which indicates that the recovery and reusing of recyclable materials were critical to achieving a high detector yield of 70\%. Further details on each step of the germanium processing and recovery follows. 
\begin{table}[htp]
\caption{\label{table:ge_load}  A summary of $^{76}$Ge-enriched material usage in the detector production process, during which thirteen zone-refined bars were produced by the detector manufacturer.  Each zone-refined bar may contain enriched materials that have only gone through the $^{76}$GeO$_{2}$ reduction process (`Virgin'); enriched materials from previously pulled crystals that could be reloaded in the zone refiner after an acid etch at the detector manufacturer (`Recycled'); and enriched materials that required additional chemical processing as described in this paper (`Recovered').  In most detector production runs, multiple detectors were fabricated from each zone-refined bar, but there were instances when the pulled crystals were found to be n-type and were not suitable for detector production. The yield of the processed virgin material at ESI and the final detector mass is relative to the initial purchase of 42.5~kg of germanium isotopically enriched  in \gess.}	
\begin{center}
\begin{tabular}{cccccc} \hline
Zone-refined & Virgin & Recycled & Recovered & No. of finished & Summed mass \\
bar number & (g) & (g) & (g) & detectors & of detectors (g) \\ \hline
 1 &	9134.3 &	0 &	0 &	3 &	3106.7 \\ 
 2 &	8812.5 &	420.1 &	0 &	2 &	2119.5  \\ 
 3 &	9168	 & 0 &	0 &	5 &	4197.1 \\ 
 4 &	0 &	9458.7 &	0 &	0 &	0 \\ 
 5 &	9218.5 &	0 &	0 &	1 &	521.1  \\ 
 6 &	5430.5 &	3618.5 &	0 &	0 &	0 \\ 
 7 &	0 &	9197 &	0 &	6 &	4349.8 \\ 
 8 &	0 &	9106.7 &	0 &	5 &	3663.4 \\ 
 9 &	0 &	8534.7 &	807.4 &	4  &	3435.5 \\ 
 10 &	 0 &	3551.1 &	4709.4 &	4 &	3846.1 \\ 
 11 &	 0 &	2557.4 &	6616	 & 2 &	2136.3 \\ 
 12 &	 0 &	5889 &	0 &	0 &	0 \\ 
 13 & 0 &	5568	 &0 &	3 &	2308.4 \\ \hline
 Total: & 41763.8 & & & 35 & 29683.9 \\ \hline \hline
 Yield: & 98.3 \% &&&& 69.8\% \\
 \end{tabular}
\end{center}
\label{default}
\end{table}%

\subsection{Enrichment and Shipping }

Isotopic enrichment was performed in the large centrifuge facility, Electrochemical Plant (ECP), in Zelenogorsk, Russia. The germanium was converted to the stable gas GeF$_4$ and introduced into a long series of centrifuges. When the required isotopic enrichment is achieved ($>$87\% \gess), the gas is bubbled into cold water and hydrolyzed. The hydrogen combines with the fluorine and the germanium with the oxygen, forming a GeO$_2$ precipitant in a dilute solution of HF acid. The HF solution is drained off and the oxide is dried to a fine powder. When not being processed, the GeO$_2$ was packed in plastic bottles, which were stored under concrete, steel, and soil to reduce exposure to cosmic-ray neutrons that cause nuclear spallation reactions that create radioactive atoms in the germanium.  The \gess\ oxide went to St. Petersburg by land, to Charleston, SC USA by ship, then to Oak Ridge, Tennessee by truck, all while enclosed a heavy iron shield described in Section \ref{sec:reduction}. These precautions reduced cosmic-ray exposure by on the order of a factor of ten.   

Upon arrival in Oak Ridge the \gess O$_2$ was randomly sampled and tested by inductively coupled plasma mass spectrometry (ICP-MS) for isotopic abundance and content of other elements. A total of 42.5~kg of enriched \gess\ in the form of 60.4 kg of GeO$_2$, were purchased from ECP. The isotopic contents of five random samples are shown in Table \ref{table:isoab}.
	
\begin{table}[t]
\caption[]{Listed are the isotopic abundances of the \MJ\ \DEM\ germanium enriched in \gess\ measured by Oak Ridge National Laboratory after acceptance of the material.  The values are given in percentage of the mass.  Samples S1 and S2 are 
from a 20-kg-shipment of Ge metal equivalent received on September 12th, 2011. Samples S3,  S4 and S5 are from the 12.5-kg-shipment of \gess\ received on October 23, 2012. Additional measurements, some by other laboratories, for a broader sampling of the \MJ\ \DEM\ material results in a weighted average of $88.1 \pm 0.7$\% \gess\ \cite{Ell16}.}
	\begin{center}
	\begin{tabular}{c|c|c|c|c|c}

			
Isotope: 	&	$^{70}$Ge	&	$^{72}$Ge	&	$^{73}$Ge	&	$^{74}$Ge	&	$^{76}$Ge	\\\hline
\hline
S1	&	$<$0.20	&	$<$0.20	&	$<$0.20	&	12.5(1)	&	86.9(9)	\\	
S2	&	0.0157(3)	&	0.0058(3)	&	0.02(1)	&	12.0(1)	&	87.9(9)	\\	
S3	&	$<$0.01	&	$<$0.01	&	0.0110(2)	&	13.06(13)	&	86.9(9)	\\	
S4	&	$<$0.01	&	$<$0.01	&	0.0345(7)	&	13.12(13)	&	86.8(9)	\\	
S5	&	$<$0.01	&	$<$0.01	&	0.0167(3)	&	12.96(13)	&	87.0(9)	\\	
										
		\end{tabular}
	\end{center}
	\label{table:isoab}
\end{table}

\begin{figure}[h]
\centering
\includegraphics
[width=0.8\textwidth]{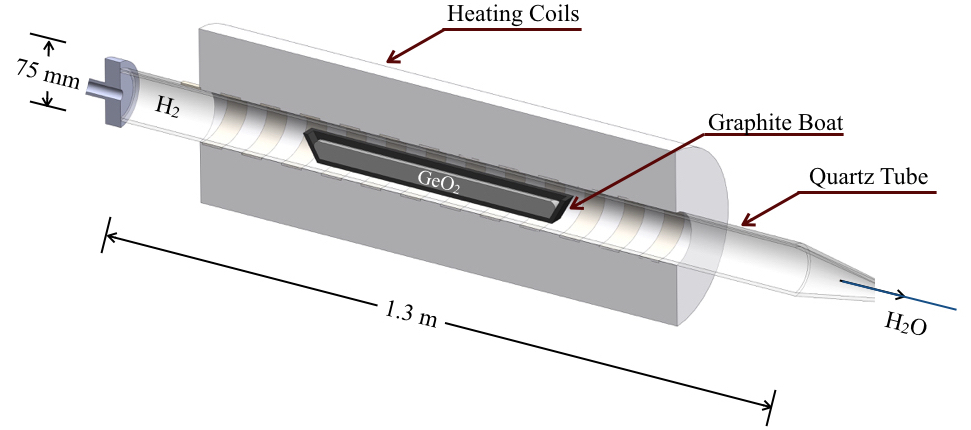}
\caption[]{A drawing of the reduction furnace including the tube that carries the water vapor resulting from the reduction reaction. When the water vapor ceases to appear, the reduction is complete.}
\label{fig:red}
\end{figure}

\subsection{Reduction of the Oxide to Metal}

The drawing of the reduction furnace depicted in Figure \ref{fig:red} consists of a cylindrical electric furnace 1.3 m long, with a bore that accommodates a 75 mm quartz tube. The \gess O$_2$ was placed in a graphite boat that holds about 800 g of oxide. The quartz tube of the furnace was capped at both ends with Teflon plugs, one with a gas supply tube and the other with a gas exit tube.  The reduction was done under a flow of pure hydrogen at 650 $^\circ$C for approximately 9 hours, thus converting the germanium oxide to metal and water vapor.  The unreacted hydrogen exited and served as a carrier gas for removing the water vapor product. The reduction produced a fine germanium metal powder and H$_2$O, which was vented. When the water ceased to appear at the vent, the reduction was complete. At that point, the hydrogen gas was flushed out with nitrogen, and the temperature was raised to 1030 $^\circ$C in the N$_2$ atmosphere until the germanium powder melted. Then the furnace temperature is slowly lowered. The production of metal was about 500 g/day per furnace. The average germanium yield of the reductions was greater than 99\%.

\subsection{Zone Refinement}

Zone refinement was developed for silicon and germanium at the Bell Laboratory in 1966 and is the standard technique used to purify germanium \cite{Pha66, Mul}. Germanium is a metal in which the liquid phase has a larger affinity for impurities (larger segregation coefficient) than the solid phase. In the zone-refinement apparatus, RF coils surround the sample and create a narrow region of liquid phase. When metal germanium in a graphite boat is moved slowly through the coil, the liquid
region moves through the metal carrying some fraction of the impurities with it. In the present case, the graphite boat was moved very slowly (1.5 mm/min) through the quartz tube surrounded by six RF loops. In the case described here, there were six coils through which the boat moved.       

\begin{figure}[h]
\centering
\includegraphics
[width=0.45\textwidth]{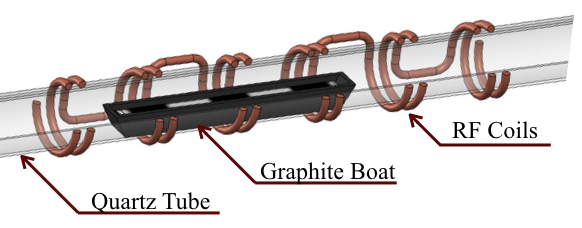}
\includegraphics
[width=0.45\textwidth]{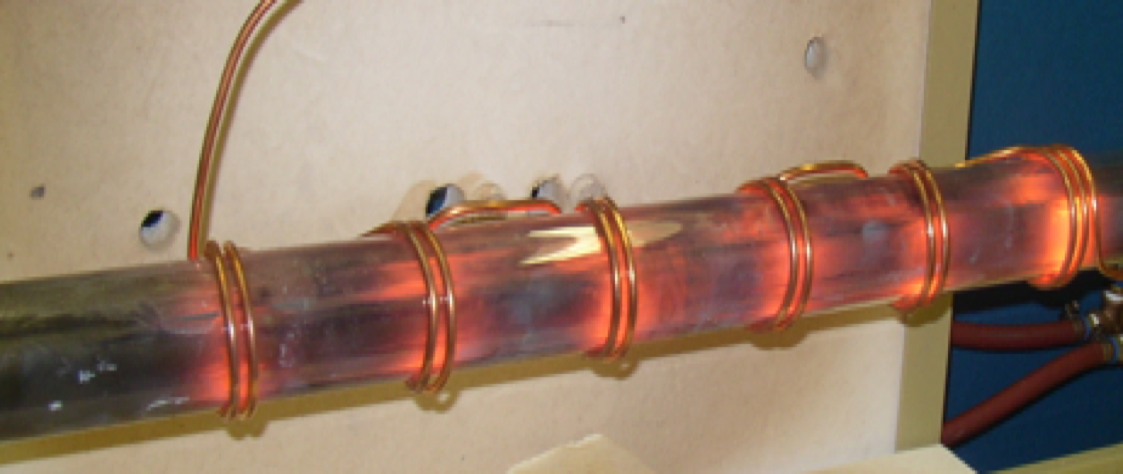}
\caption[]{A (left) drawing and (right) photograph of the six-coil array is shown during operation with a germanium-loaded graphite boat  moving slowly through the series of RF coils.  The bright regions in the drawing depict the regions of melted germanium. }
\label{fig:boat}
\end{figure}

The power was adjusted to create a narrow liquified region inside each coil. These liquefied regions were 2 -- 3 cm in length. The germanium metal was zone refined to a resistivity of $47\ \Omega\cdot$cm as required by detector manufacturers. The RF coils are shown in Figure \ref{fig:boat}.

The fraction of the zone-refined bar that met the required $47\ \Omega\cdot$cm resistivity after the first zoning was between 65\% -- 80\% of the length. The higher the purity of the input metal from the reduction furnace, the higher the fraction of the zone-refined bar that met the required resistivity. In the Ge-detector industry, the usual practice is to send the tails that do not meet that standard to chemical reprocessing (see Section \ref{sec:reprocessing}), which has an average overall yield of 70\%. In the case of enriched germanium those losses were avoided by continuously zone refining the tails until the fraction of the bars not meeting the required standard was small, between 15 -- 25\%. Only a tiny fraction of the material is lost in the zone refining process, while losses between 30\% and 35\% or more are common in chemical reprocessing. This practice allowed the project to achieve an overall yield of 98.3\% of virgin enriched germanium meeting the $47\ \Omega\cdot$cm resistivity throughout the first cycle from oxide through the zone-refinement.  

\subsection{Reprocessing of the Scrap Germanium and Acid Etch}\label{sec:reprocessing}

The term ``reprocessing'' includes repeated zone refining, chemical processing of metal scraps, chemical processing of the sludge formed from cuttings and grindings in a lubricant bath, and chemical reprocessing of the acid-etch solutions in order to recover germanium metal. The \MJ\ \DEM\ program recovered germanium through reprocessing for all material except for the chemical reprocessing of acid-etch solutions, for which there was instead a feasibility study undertaken in this project. Losses were minimized by additional zone refining of the portion of the bars that did not meet the $47\ \Omega\cdot$cm standard. This was repeated until most of the remaining bar did not meet the standard, and continuing zone refining was no longer practical. At that point, the remaining tails were chlorinated, condensed and collected as GeCl$_4$, which after fractional distillation, was hydrolyzed in cold water to form a GeO$_2$ precipitate in a solution of HCl. The GeO$_2$ is dried and reduced to metal, as was the original oxide. Following reduction, the bars are zone-refined. 

The material referred to as ``sludge'' is a combination of metal grit in a water and lubricant solution used in the machining and grinding of the detector blank, and requires chemical reprocessing. It was found to be most efficient to pour the sludge into large area tubs and wait for the solid material that contains the germanium to settle to the bottom. The liquid on top was tested with an atomic absorption spectrometer to determine the germanium content. After about a week to ten days, the germanium content in the remaining water solution was minimal. The liquid was then poured off and the solid material was chlorinated, hydrolyzed, reduced, zone-refined and sent to the detector manufacturer to make additional detectors.

Finally, a feasibility study was undertaken to recover germanium from the typical acid-etch solution used in detector fabrication. The contents of the solution by volume were: HF (2\%), HNO$_3$ (9.5\%), Methanol (54.5\%) and H$_2$O (34\%). Atomic absorption analysis determined that 19 liters of solution contained  an average of 80 g of germanium. The content of HF acid negated the use of glassware. Ion exchange was eliminated because of the probable interaction of HF acid with the ion-exchange medium. Accordingly, distillation using Teflon components was considered the most practical solution. However, Teflon has a maximum practical operations temperature of about 2000 $^\circ$C. In addition, Teflon has a very low thermal conductivity. Nevertheless, Teflon components can be used in a microwave boiler tuned for water to make a practical high-volume apparatus. Following distillation, 190 g of germanium salts were recovered from 19 liters of etch solution. 
While the final chemistry to recover germanium from the salts was not completed, the estimate is that 90\% recovery of germanium from the etch solution is achievable. This would be very important for a large, possibly ton-scale, \zBB -decay experiment with enriched germanium. 

\section{Generation of Internal Radioactivity in the Germanium by Cosmic-Ray Neutrons}\label{sec:cosmo}

Energetic neutrons from cosmic rays at the earth's surface form radioactive isotopes in the germanium itself, which can produce serious background to \zBB -decay experiments. Extensive studies have been conducted to determine the generation of radioactive isotopes produced by energetic neutrons on \gess\ \cite{Avi92, Ell10}. The following isotopes (half lives in parentheses) have been observed and measured: $^{54}$Mn(312 d), $^{57}$Co(272 d), $^{65}$Zn(244 d), $^{68}$Ge(271 d), $^{60}$Co(5.28 y) and $^3$H(12.3 y). Both $^{60}$Co and $^{68}$Ge contribute background in the \zBB\ decay region of interest. All but one of these isotopes are essentially eliminated by zone refining and by the Czochralski crystal pulling operation. However, $^{68}$Ge is unaffected by these processes, but is essentially eliminated by the initial isotopic enrichment of \gess\ by centrifugation. The isotope $^{68}$Ge has a half-life of 271 days, and decays to $^{68}$Ga, which decays by electron capture to $^{68}$Zn, followed by gamma de-excitation with a total energy release of 2921 keV. This is well above the \qval = 2039.061(7) keV \cite{Mou10} and can directly interfere with the search for the \zBB\ decay of \gess. This background is minimized by keeping the material well shielded from cosmic rays whenever possible. 

\section{Reduction of Cosmogenic Activation During Shipping from Russia}\label{sec:reduction}

\begin{figure}[h]
\centering
\includegraphics
[width=0.8\textwidth]{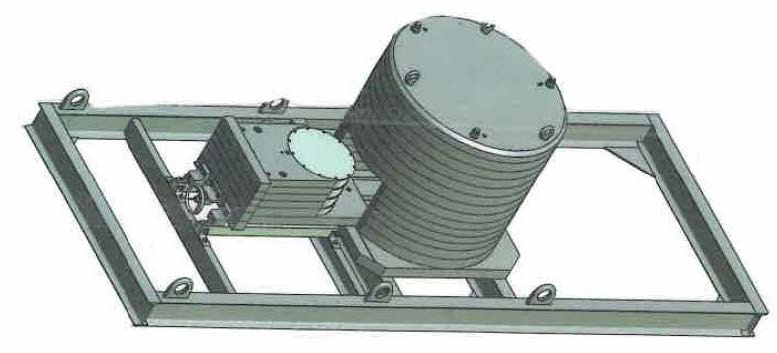}
\caption[]{A computer-aided design drawing of the Russian shipping shield.  The construction is of steel plates.  There is 72 cm of steel above the cylindrical chamber containing the GeO$_2$ and 43 cm on the sides.  The small, light-colored cylinder on the pull-out drawer is the GeO$_2$ container.}
\label{fig:ship}
\end{figure}

To minimize cosmogenic production of $^{68}$Ge, protection from cosmic-ray neutrons must start as soon as possible after the GeF$_4$ gas leaves the centrifuge. From the centrifuge, the gas is hydrolyzed, and the resulting GeO$_2$ is extracted, dried, and stored in plastic bottles. At that point it is placed under concrete, steel, and soil overburden to reduce activation. Shipment of the material at the lowest elevation possible minimizes the cosmic-ray exposure.  The first shipment of GeO$_2$ in September 2011 was taken by truck in a special steel transport shield to the port of St. Petersburg where it went by ship to Charleston, South Carolina. After 5 days in U.S. customs, the shipment went by truck to Oak Ridge, Tennessee. During its travel from Zelenogorsk, Russia to Oak Ridge, the material always resided in a steel shield designed and built in Russia to reduce activation of $^{68}$Ge. A drawing of the shield is shown in Figure \ref{fig:ship}. Calculations by Barabanov et al. \cite{Bar06}, show the container reduced the cosmogenic generation of $^{68}$Ge by a factor of approximately ten (see Table \ref{table:crexp}).

While the steel shipping container was effective in reducing cosmogenic production of radioactive backgrounds in the germanium during transportation, other shielding measures were necessary during and after the processing described above. The germanium, in various forms, was kept under an overburden during all times when the GeO$_2$ and Ge metal were not actively being processed. Each zone refining cycle took approximately 14 hours and ended in the early morning hours. The material was cooled and brought underground within two hours after completion. Furthermore, the germanium was stored underground whenever possible during detector fabrication. To achieve the underground storage requirement, space was rented at the bottom of Cherokee Caverns located only a few miles from the germanium-processing site in Oak Ridge, TN. The cosmic-ray muon flux was measured by Y. Efremenko et al. \cite{EfrComm}, and it was determined that the production of the isotope $^{68}$Ge was reduced by a factor of approximately ten from that of unshielded germanium.  Shielding of the GeO$_2$ at Zelenogorsk and shipment in the shield, combined with minimizing the time above ground during reduction and zone refinement at the germanium-processing site and during detector fabrication, resulted in a significant reduction in radioactive isotopes in the bulk germanium. 
Estimates of the times and exposure reductions of all of these steps in the process are shown in Table \ref{table:crexp}. Shipment 2 consisting of 17.8 kg of GeO$_2$, containing 12.5 kg of Ge, arrived in Oak Ridge on 23 October 2012. Very similar procedures for protection from cosmic-ray exposure were used  with very close to the same exposures at each step of the production, shielding,  and durations for the various steps of the transportation.

\begin{sidewaystable}[th]
\caption[]{Reduction factors, the ratios of the calculated unshielded to shielded exposures computed by Monte Carlo simulations supplied by the \Gerda\ Collaboration. Detailed reduction factors for all 5 stable Ge isotopes are given in reference \cite{Bar06}.}
	\begin{center}
	\begin{tabular}{c|c|c|c|c|c}

			Event Time		&	Event				&	Event 		&	Shielding 		&	Reduction 		&	Comments			\\
			 and Date			&						&	Duration		&	Description	&	Factor			&						\\\hline
			7:30 AM			&	Complete 				&	189 days		&	Underground	&	22 estimated		&	Concrete, dirt			\\
			06/28/2011		&	enrichment of 			&				&	Storage/ECP	&	by ECP 			&	and steel shielding		\\
							&	shipment 1			&				&				&	personnel			&	await steel container		\\\hline	
			08/02/2011		&	Load into 				&	1 hour,		&	No shielding	&	10.4$\pm$0.9		&						\\	
							&	shipping container		&	32 min		&				&					&						\\\hline								
			08/02/2011		&	Start ground		 	&	8 days		&	Steel shipping	&	10.4$\pm$0.9		&						\\
							&	transport to			&				&	container		&					&						\\
							&	St. Petersburg			&				&				&					&						\\\hline
			08/12/2011		&	Depart 				&	25 days		&	Steel shipping	&	10.4$\pm$0.9		&						\\								
							&	St. Petersburg			&	crossing		&	container		&					&						\\\hline
			09/07/2011		&	Arrive Charleston,		&	5 days		&	Steel shipping	&	10.4$\pm$0.9		&	Exact time container		\\								
							&	unload, clear			&				&	container		&					&	removed from ship		\\
							&	customs, truck to		&				&				&					&	unknown.				\\
							&	Oak Ridge			&				&				&					&						\\\hline
			8:00 AM - 11:30 AM	&	Unload, weight			&	3.5 hours		&	None		&	0.0				&						\\								
			09/07/2011		&	samples, place			&				&				&					&						\\
							&	in cave				&				&				&					&						\\
							&						&				&				&					&						\\\hline
			11:30 AM			&	Placed GeO$_2$		&	Remains in	&	Underground	&	$\backsim$10		&	$\backsim$ 12.5 days	\\													
			09/12/2011		&	in cave storage			&	cave when 	&	130 ft of rock	&	an estimate based	&	Total effective exposure  	\\											
							&						&	not being		&				&	on muon reduction	&	from 6/28/2011 until		\\
							&						&	processed		&				&	factor			&	placed in the cave		\\\hline							
		\end{tabular}
	\end{center}
	\label{table:crexp}
\end{sidewaystable}

The average overall estimated sea-level equivalent exposure for all detectors, excluding detector manufacturing, was 12.5 days. 
These steps, combined with the storage underground in the Cherokee Caverns between processing steps, greatly reduced the overall cosmic-ray activation production of radioactive isotopes. The exact reductions for the individual isotopes are not known at this time, however, the reductions are being measured by the collaboration at the time of this writing. 

\begin{figure}[h]
\centering
\includegraphics
[width=0.9\textwidth]{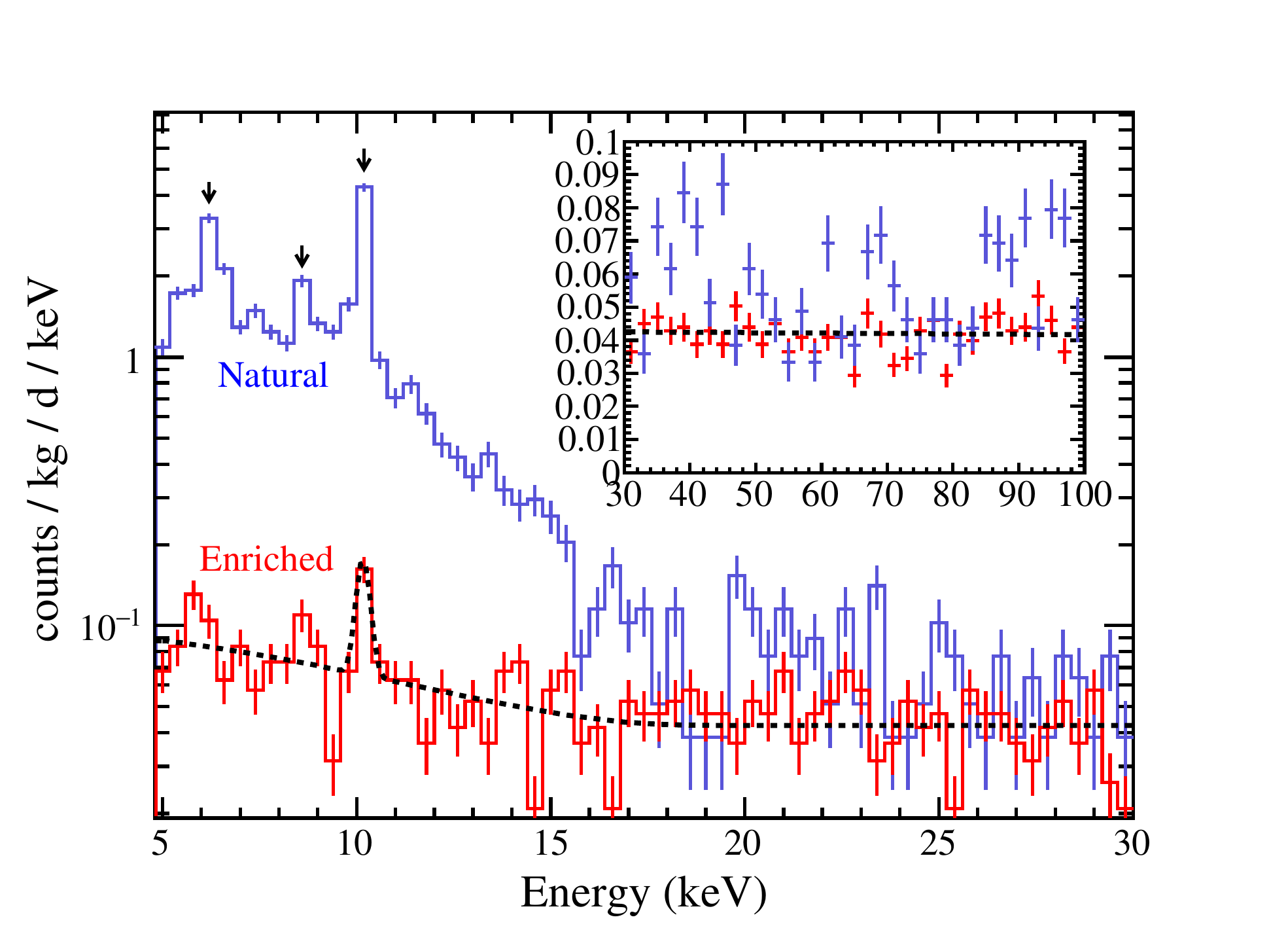}
\caption[]{Energy spectra from 195 kg d of natural
(blue) and 478 kg d of enriched (red) detector data. A fit
of the background model (linear + tritium beta spectrum +
$^{68}$Ge K-shell) to the enriched detector spectrum is also shown (dotted
black). The spectrum demonstrates that the 10.4-keV X-ray line from the electron capture of $^{68}$Ge to $^{68}$Ga is reduced by a factor of 30 between the natural and enriched detectors. In addition, the continuum of the blue curve is most likely dominated by tritium decay. The other prominent X-rays peaks are dominated by the electron capture decay of $^{65}$Zn (8.9 keV) and $^{55}$Fe (6.5 keV). Figure originates from Ref. \cite{Abg16}.}
\label{fig:spec}
\end{figure}

The final effect of this program's cosmogenic reduction in the enriched material is evident from the early data from the \MJ\ \DEM\ shown in Figure 6, in which both curves are normalized to the same detector exposure in counts/kg/d/keV. 
For comparison, the detectors labeled ``natural'' were fabricated from natural abundance germanium, which was not shielded and was saturated by radioactive isotope production from energetic cosmogenic neutrons. The large continuum is largely due to the decay of cosmogenically produced tritium. In the natural detector data, there is a large X-ray peak near 10.4 keV from the electron capture of $^{68}$Ge to $^{68}$Ga and another near 8.9 keV from the electron capture of $^{65}$Zn. The X-ray peak near 6.5 keV is likely a mixture of X-rays from the electron capture decays of $^{55}$Fe, $^{54}$Mn, and $^{57}$Co. In addition to the difference in cross sections for cosmogenically activated isotopes between natural and enriched germanium, it is very clear that the precautions taken from the time of the enrichment in Russia, all the way to Oak Ridge, Tennessee, as well as the precautions taken during the processing, reprocessing, and fabrication of enriched detectors, has been very successful in reducing the internal background from cosmic-ray induced radioactivity in the enriched germanium. 
More data will be needed before the total background in the region of interest can be determined, which is the focus of a separate study and an upcoming publication on the cosmogenic backgrounds measured in the \MJ\ \DEM. Nonetheless, Fig.  \ref{fig:spec} shows a significant reduction in all of the low-energy X-ray peaks. The factor of 30 reduction in the Ga X-ray peak from the electron capture decay of $^{68}$Ge implies a similar reduction in backgrounds at higher energies due to the decays of $^{68}$Ge and $^{60}$Co. 




\section{Discussion and Conclusions}

The purpose of designing, building and operating the \MJ\ \DEM\ is to test the technology and method of using point-contact germanium detectors fabricated from germanium enriched in the \BB-decay candidate isotope \gess, mounted in copper cryostats electroformed underground, for a ton-scale \gess\ \zBB-decay experiment. This goal requires that the background be reduced to the lowest practical level and that the fraction of the expensive enriched germanium that ends up in high-quality detectors be maximized. 

To demonstrate this, the \MJ\ Collaboration purchased GeO$_2$ containing 42.5~kg of germanium isotopically enriched from 7.83\% to 88\% in \gess. A special facility was set up by the Collaboration in Oak Ridge, Tennessee, and managed by Electrochemical Systems, Inc. Equipment was purchased and fabricated to equip the facility to reduce the oxide to metal, to zone refine the metal to a resistivity of at least $47\ \Omega\cdot$cm, to test the metal and to reprocess the scrap from the detector manufacturer. The total yield in the fraction of mass of germanium that ended up as detector-grade germanium through the processing of virgin material was 98.3\%. Finally, thirty-five point-contact \gess-enriched detectors having a total mass of 29.7 kg were fabricated for the \MJ\ \DEM.  This represented an overall yield of detector mass to that of purchased material of approximately 70\%, with 2.64 kg of scrap germanium remaining. 
After careful evaluation, it was discovered that the largest loss was from  machining and etching the germanium at various stages of the detector manufacturing process. While the mixed HF and HNO$_3$ acid-etch solution was not reprocessed, a detailed protocol was developed to accomplish this for the first time. An experimental determination was made that the yield for extraction of germanium from the acid mixture might be as high as 90\%. We estimate that this additional reprocessing step would likely increase the overall yield in the mass of operating detectors to between 80\% and 85\%. The estimated 15\% to 20\% irrecoverable losses would come mainly from the chemical reprocessing. However, further R\&D could in principal recover much of those losses. Our estimate of the costs of these steps, however, is that they might be equal to or greater than the value of the recovered germanium.

Procedures were developed to minimize the cosmic-ray activation of radioactive isotopes in the enriched germanium. These measures involved the storage of the newly enriched GeO$_2$ under an overburden immediately after enrichment, the fabrication of a special steel shipping container, and underground storage near Oak Ridge at all times when the material was not being processed. The results of these efforts are reflected in Figure \ref{fig:spec}. Finally, the very low background in the low-energy region shown in Figure \ref{fig:spec} implies that the \MJ\ \DEM\ will be very effective in the search for cold dark matter and for axions generated by low energy atomic processes in the sun, the fluxes of which were calculated by Redondo \cite{Red13}.

An important issue in the consideration of a ton-scale, isotopically enriched \gess\ \zBB-decay experiment is the cost. The high cost of the isotope requires the highest possible yield in detector production. A summary of the material yields at various stages of the processing is as follows:
\begin{itemize}
\item Total yield of processing virgin GeO$_2$:  98.3\%
\item Average yield of first zone refining: 65\% - 75\%
\item Average yield of chemical reprocessing: 70\%
\item Final overall yield of  \MJ\ \DEM\ detector mass per purchased mass of Ge: 69.8\%
\item Potential yield of detector mass per purchased mass of Ge 85\%
\end{itemize}
These values and the R\&D work done by the collaboration clearly demonstrate that the experience gained in the processing of the germanium for the \MJ\ \DEM\ justifies the conclusion that the yield in detectors could possibly be as large as 85\%. 
The production rate needed for a ton-scale \gess\ \zBB-decay experiment would require four reduction furnaces of the type used for the \MJ\ \DEM\   and two zone-refinement apparatuses.
The only new facility required would be a special chemistry laboratory for the chemical processing of the acid-etch solution. The feasibility of efficiently processing enriched germanium for a ton-scale experiment has been  established by the work described in this article.

\section*{Acknowledgements}
This material is based upon work supported by the U.S. Department of Energy, Office of Science, the Office of Nuclear Physics under Award Numbers DEAC02-05CH11231, DE-AC52-06NA25396, DE-FG02-97ER41041, DE-FG02-97ER41033, DE-FG02-97ER41041, DE-FG02-97ER41042, DE-SC0012612, DE-FG02-10ER41715, DE-SC0010254, and DE-FG02-97ER41020. We acknowledge support from the Particle Astrophysics and Nuclear Physics Programs of the National Science Foundation through grant numbers PHY-0919270, PHY-1003940, PHY-0855314, PHY-1202950, MRI 0923142, PHY-1307204 and PHY-1003399. We acknowledge support from the Russian Foundation for Basic Research, grant No. 15-02-02919. We acknowledge the support of the U.S. Department of Energy through the LANL/LDRD Program.  

\section*{References}

\bibliography{Ge-processing}

\end{document}